\documentclass[11pt,tightenlines]{revtex4-2}

\usepackage{graphicx}
\usepackage{caption, subcaption} 
\usepackage{sidecap}
\usepackage{dcolumn}
\usepackage{amsmath}
\usepackage{amsfonts}
\usepackage{gensymb}
\usepackage{bm}
\usepackage{float}
\usepackage{xcolor}
\usepackage[normalem]{ulem}
\begin{document}
\title{Critical transport behavior in quantum dot solids}

\author{Zachary Crawford$^{1}$\footnote{These authors contributed equally to this work}, Adam Goga$^{1}$\footnotemark[1], Mikael A. Kovtun$^1$, and Gergely T. Zimanyi$^1$}
\affiliation{$^1$Physics Department, University of California, Davis, Davis CA 95616}

\date{\today}

\begin{abstract}
Due to recent advances, silicon solar cells are rapidly approaching the Shockley-Queisser limit of 33\% efficiency. Quantum Dot (QD) solar cells have the potential to surpass this limit and enable a new generation of photovoltaic technologies beyond the capabilities of any existing solar energy modalities. The creation of the first epitaxially-fused quantum dot solids showing broad phase coherence and metallicity necessary for solar implementation has not yet been achieved, and the metal-insulator transition in these materials needs to be explored. We have created a new model of electron transport through QD solids, informed by 3D-tomography of QD solid samples, which considers disorder in both the on-site and hopping terms of the commonly studied Anderson Hamiltonian. We used the transfer matrix method and finite-size scaling to create a dynamic metal-insulator transition phase diagram. For a surprisingly large portion of the parameter space, our model shows a critical exponent distinct from the expected value for the Anderson transition. We show the existence of a crossover region from the universality class of the Anderson transition (AI) to the Chiral Orthogonal class (BDI) due to the addition of weak kinetic (hopping) disorder.
\end{abstract}

\maketitle
\section{Introduction}

Colloidal semiconductor quantum dots (QDs) are well-defined nanoscale building blocks of mesoscale materials that can be fabricated with increasingly consistent control of size, composition, and shape. The energetics and interactions (electronic, excitonic, magnonic) in QD solids are tunable by changing the QD size, size distribution, shape, inter-QD spacing, spatial ordering, surface chemistry and defects, and the properties of the matrix between the QDs \cite{Liu2010,xiaolei,an_electron_2007,kang_electronic_1997,Scheele2013}. This remarkable tunability makes QD superlattice (SL) solids exciting materials for improving optoelectronic applications\cite{talapin_prospects_2010,doi:10.1021/nn506223h} including third generation solar cells \cite{Nozik02,doi:10.1021/jp806791s}, light emitting diodes \cite{shirasaki2013emergence}, and field effect transistors (FETs)  \cite{Talapin07102005,hetsch_quantum_2013}. QD SL solids are well suited for photovoltaic (PV) applications because the band gap can be tuned via QD radius to improve PV power conversion efficiency. Cells constructed from QD have the opportunity for efficiencies beyond the Shockley-Queisser limit of $\sim$33\% \cite{Shockley1961} which restricts conventional silicon solar cells.

In recent years, quantum dot photovoltaics (QD-PV) had the fastest rising efficiency of any PV, having broken the 10\% level around 2015, and reaching 18\% in 2021 \cite{Hao2020,Sanehira_13.4percent}. The key factors controlling this efficiency are absorption and transport. One essential factor limiting the utility of QD optoelectronics is the high spatial and energetic disorder in QD solids, though current research is actively improving control of this disorder during fabrication\cite{whitham,evers,Diroll2015,Altamura2012,xiaolei,Liu_2013,Abelson_2019}. This type of disorder in QD SL causes decoherence in the electronic wave functions between individual QDs which results in weakly-coupled QD materials with slow hopping transport. 

The industry workhorse PbSe QDs have highly localized electron wavefunctions, and thus layers made of these PbSe QDs are in the insulating phase with carrier mobilities orders of magnitude lower than in c-Si \cite{kang_electronic_1997,Liu_2013}. The most promising direction to improve transport is to drive these QD solids through the Metal Insulator Transition (MIT). Metallicity can be reached by establishing phase coherence and mini-band formation \cite{Lazarenkova2001,klos2008electronic}. The field of MITs is decades-old and extensive, however most studies focus on simplified, paradigmatic models rather than examining modern experimental capabilities. Techniques for the fabrication and characterization of epitaxially-fused QD solids are rapidly advancing and it is necessary for modeling efforts to follow suit \cite{Abelson_2019, xiaolei}. In particular, projects by Law and Moulé have generated high-quality 7-layer QD solids, which have been imaged and characterized using 3D near-atomistic tomography. They have determined detailed features of individual QDs, as well as couplings between the QDs \cite{xiaolei}. Their samples are approaching metallicity, but so far do not show evidence of crossing the MIT. Thus, this work charts a road map for the fabrication processes and parameters to drive the QD solids through the elusive MIT. 
\par
To achieve this goal, we have developed a comprehensive simulation of the transport properties of QD solids that captures both the disorder in size and couplings. There have been multiple numerical studies exploring the MIT with either exclusively on-site disorder, or kinetic (hopping) disorder \cite{MacKinnon1983,Slevin2014,Slevin2004}. However, real materials exhibit disorder of both types. Reducing disorder by increasing the uniformity and quality of both QDs and their epitaxial couplings will be key to driving these QD solids across the MIT, and thereby creating the first QD solid showing high enough electron mobility for competitive photovoltaic usage.
\section{Model}
\par
We created a model of electron transport in QD superlattice (SL) solids with the necessary features to approach and capture the metal-insulator transition from the metallic side. Through experimental testing of QD SLs, we found that the disorder in both the size of the QDs and the distance between QDs dominates the resulting electronic properties of the QD SL film \cite{kang_electronic_1997}. Our model followed this guidance with a tight-binding Hamiltonian (equation \ref{eq:model}) which has disorder in the on-site energy term ($\epsilon_i$), as well as ``kinetic disorder" in the nearest-neighbor hopping term ($t_{ij}$). The on-site disorder accounts for the size disorder, and therefore energetic disorder, in the individual QDs \cite{LumanPerc}. Kinetic disorder in the hopping term represents the non-uniform distribution of experimentally-observed epitaxial electronic couplings between QDs. 

\begin{equation}\label{eq:model}
H=\sum_i \varepsilon_i c_i^\dagger c_i + \sum_{<i,j>}t_{ij} c_j^\dagger c_i + \mathrm{h.c.}
\end{equation}

\begin{figure}[htb!]
    \centering
    \includegraphics[width=0.75\linewidth]{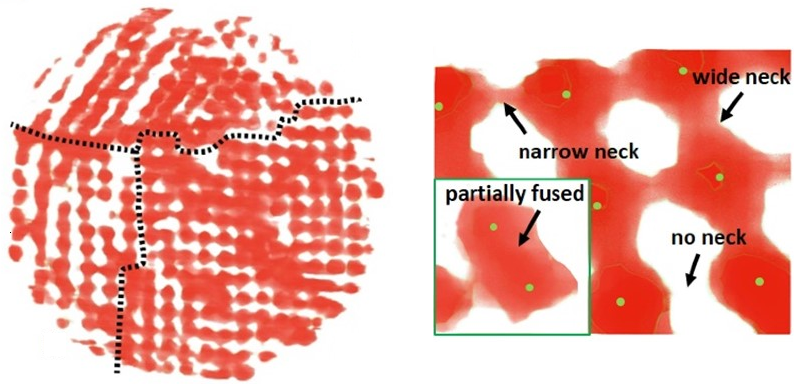}
    \caption{(a) A slice of the tomogram of a 7-layer QD SL solid sample, showing in plane necks between the QDs. (b) Magnified iso-surface views of two regions of the sample to illustrate typical neck polydispersity and a highly-fused pair of QDs(inset). The green dots denote the center-of-mass of each QD. Reproduced from Chu et al. \cite{xiaolei}.}
    \label{fig:tomography}
\end{figure}

Our studies investigated quasi-1D carrier transport in 3D systems, where sufficient on-site disorder is known to drive the system through the MIT in accordance with traditional Anderson localization theory. Therefore, one of our primary motivations in this study was to address how the inclusion the effect of kinetic disorder affects both critical behavior and the underlying universality class. We used periodic boundary conditions in the transverse direction to connect nearest-neighbor sites, and drew $\varepsilon_i$ from a box distribution to represent variations in QD size and quality [$-W/2, W/2$]. Our implementation of disorder in the hopping integral $t_{ij}$ is inspired by the samples created by Law and characterized by Moulé, which have shown epitaxial couplings between up to 72\% of the QD-QD pairs \cite{xiaolei}. The epitaxial couplings induce a much-enhanced electronic coupling between QD-QD pairs, i.e. when QD pairs have an epitaxial neck, the propensity of an electron to hop is severely increased. These epitaxial couplings are shown in figure \ref{fig:tomography}. Thus, the kinetic disorder can be best captured by a binary distribution and $t_{ij}$ is drawn with probability $c$ to be $t_{H}$ to represent epitaxially-coupled QD pairs, and with probability $(1-c)$ to be $t_{L}$ to represent QD pairs without epitaxial coupling. To fix the energy scale, we set $t_{H}=1$. We note that while traditional tight-binding Hamiltonians use a negative hopping term, the formulation in equation \ref{eq:model} becomes equivalent in the transfer matrix implementation due to the on-site disorder's symmetry about $W=0$.

\begin{figure}[hbt!]
    \centering
    \includegraphics[width=0.75\linewidth]{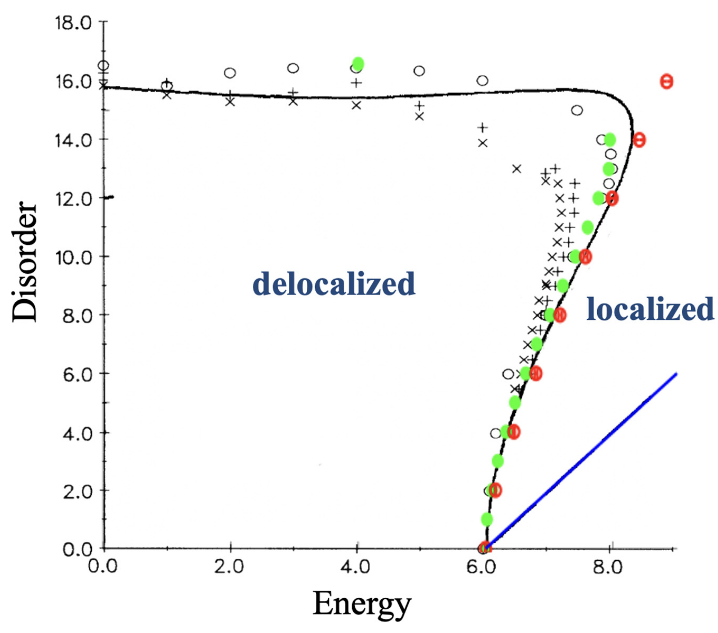}
    \caption{Phase diagram of localization in the (energy, disorder) plane for a uniform disorder. Due to the $E\rightarrow-E$ symmetry of the problem, only the positive part ($E\geq0$) of the diagram is represented. The straight blue line to the right represents the edge of the spectrum. The black crosses and circles correspond to simulation data from Grussbach and Schreiber \cite{Grussbach1995}, while the red circles show the percolation threshold of the localization landscape effective potential and the green filled circles show the computations performed by Filoche et al. of the mobility edge using the transfer matrix method. The continuous black line is the prediction of the self-consistent theory \cite{Kroha1990}. Reproduced from Filoche et al. \cite{Filoche2024}}
    \label{fig:WE-phase}
\end{figure}

We explored the MIT in the ($W$, $c$, $E$) space, and used values of $W$ such that $W/t_{H}<W_c/t<W/t_{L}$, to prevent our system from approaching the traditional Anderson localization limits. Traditional Anderson localization depends both on the diagonal disorder $W$ and the distance from the Fermi level $E$, with limits such as $W_c/t=16.53$ is the known critical value for the MIT in the purely diagonally-disordered Anderson model with a constant hopping integral $t_{ij}=1$ at the Fermi level \cite{Slevin2014}. The entire phase boundary in ($E,W$) parameter space is more complicated, with potential for reentrant behavior, and is shown in figure \ref{fig:WE-phase} as reproduced from work by Filoche et al \cite{Filoche2024}. We drove the system across the MIT by keeping $W$ fixed and varying $c$, the fraction of $t_H$ couplings. The system crosses the MIT when the concentration of $t_H$ couplings crosses a critical concentration, i.e. when $c > c_c$. We have set $t_{L}=0.3$ throughout this work to maintain the possibility of hopping in any QD-QD pair, regardless of coupling. 

\subsection{Transfer Matrix Method}

We adapted the transfer matrix method (TMM) and finite-size scaling (FSS) methods developed by Slevin et al. to determine the critical necking fraction $c_{c}$ and the critical exponent $\nu$ corresponding to the divergence of the localization length \cite{MacKinnon1983,Slevin2014,Slevin2004}. We considered the transmission of electrons with an energy $E$ relative to the Fermi level ($E=0$) through long disordered slabs with a uniform square cross section: $L_z\gg L_x=L_y=L$. We used periodic boundary conditions in the transverse ($x,y$) directions to connect slab edges and prevent edge effects. The transfer matrix method begins with a reformulation of the Schrodinger equation:

\begin{equation}
H_{n,n+1} A_{n+1}=(E-H_n)A_n - H_{n,n-1}A_{n-1}
\end{equation}
where $A_n$ is a vector representing the wave function amplitudes at sites $i$ in the $n$th slice of the wire. $H_n$ is the intra-slice Hamiltonian and $H_{n,n\pm1}$ is the inter-slice Hamiltonian. This was rearranged into a transfer matrix:  
\begin{equation}\label{eq:TMamp}
    \begin{pmatrix}
    A_{n+1}\\
    A_n\\
    \end{pmatrix}
    =
    T_n
    \begin{pmatrix}
     A_n \\
     A_{n-1}\\ 
    \end{pmatrix}
\end{equation}
where the transfer matrix is defined by:	
\begin{equation}
    T_n=
    \begin{pmatrix}
    H^{-1}_{n,n+1}(E\mathbb{I}-H_n) & -H^{-1}_{n,n+1}H_{n,n-1} \\
    \mathbb{I} & 0\\
    \end{pmatrix}
\end{equation}
The $2L\times2L$ transfer matrix propagated the wavefunction amplitudes from slice-to-slice. By iterating equation \ref{eq:TMamp} recursively, it was possible to calculate the amplitudes as they propagated longitudinally to the end of the slab. As the on-site disorder was increased, the system crossed from its initially metallic phase across the MIT into the (Anderson) localized phase. Here, the amplitude of the wavefunction decayed exponentially with a quasi-one-dimensional localization length $\xi$, which the method captures in the $L_z\gg\xi$ regime. The TMM estimated $\xi$ by computing the Lyapunov exponents of the product matrix $M$, generated by multiplying the transfer matrix $L_z$ times \cite{Slevin2004}. The Lyapunov exponents are defined as:
\begin{equation}
    \gamma_i=\lim_{L_z\to\infty} \frac{\nu_i}{2L_z} 
\end{equation}
where $\nu_i$ are the eigenvalues of $\ln(MM^T)$. As we kept multiplying by the transfer matrix, the eigenvalues of interest began to exponentially shrink towards numerical precision limits, and thus we performed a QR decomposition at regular intervals, obtaining the eigenvalues in the process. The QR decomposition process involved factorizing the product matrix into an orthogonal ($Q$) matrix and upper triangular ($R$) matrix, the latter of which yields the Lyapunov exponents via its eigenvalues. The Lyapunov exponents were estimated by truncating the above equation at a large but finite length $L_z$.

Conventional TMM methods rely on single-slab self-averaging, continually performing transfer matrix multiplications until a desired convergence threshold is reached. We considered this approach in our initial investigations, and in practice we reached convergence at $L_z\sim10^6\text{--}10^7$ for transverse $L\sim8\text{--}24$, in agreement with previous studies \cite{Slevin2004,Slevin2014}. However, with our added disorder (kinetic disorder) we wanted to investigate larger transverse $L$ sizes, so we performed calculations with fixed $L_z = 10^5$ which were averaged over multiple realizations of disorder (DR) instead of relying on self-averaging in a single slab. For most of our parameter space, we averaged across $20-45$ DR. We chose to lower the longitudinal length $L_z$ to capture the increased disorder in the transverse direction within achievable computational limits. We performed all TMM calculations on the University of California, Davis Farm computing cluster. All realizations of disorder were averaged with standard deviation weighting to account for uncertainty propagation. Finally, the quasi-one-dimensional localization length was found using the smallest positive Lyapunov exponent $\gamma_+$ via $\xi=1/\gamma_+$.

\subsection{Finite-size scaling}\label{sec:fss}
 
The single-parameter finite-size scaling method\cite{MacKinnon1983,Kramer_2010,Slevin2014} was used to extract information about critical phenomena of the system. Using data for $\xi$ over a wide range of system sizes $8 \leq L \leq 30$, and $c$ values near critical kinetic disorder, we were able to estimate universal properties of interest such as the critical exponent and critical normalized localization length. 

First, we assumed that the disorder and system size dependence on the dimensionless quantity $\Gamma=L/\xi$ are described by a scaling function of the following form \cite{Slevin_1999,Slevin2014}:

\begin{equation}
\Gamma=F(\chi L^{1/\nu},\Psi L^y)
\end{equation}

where $\chi / \Psi$ are the relevant/irrelevant scaling variables. The scaling function $F$ was Taylor-expanded in both relevant and irrelevant scaling variables up to orders $n_R$ and $n_I$ respectively. The scaling variables $\chi(c)$ and $\Psi(c)$ were Taylor-expanded as well to account for nonlinearities, up to orders $m_R$, $m_I$ respectively, shown in equation \ref{eq:taylorex}. The contribution of the irrelevant scaling variable vanished in the large L limit, such that the exponent $y$ turned negative, and $\nu$ remained positive. Our analysis was performed using $n_R=3$, $n_I=1$, $m_R=2$, $m_I=1$, though we did thorough testing of other expansion orders. Higher orders of expansion yielded diminishing returns with increased computational time, and the previously listed orders of expansion were found to be sufficient.

\begin{equation}\label{eq:taylorex}
    \chi(t)= \sum^{m_R}_{n=1}b_n t^n ,\qquad \psi(t)=\sum^{m_I}_{n=0}c_n t^n
\end{equation}

\begin{figure}[!htb]
    \centering
    \includegraphics[width=0.75\linewidth]{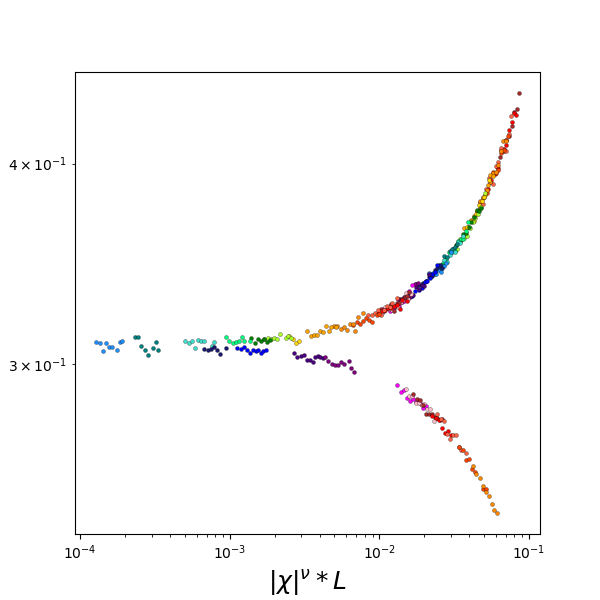}
    \caption{A log-linear plot showing the values of $\Gamma=L/\xi$ produced by our finite-size scaling algorithm for $E=2$, and $W=10$ using $17 \leq L \leq 30$ with $\nu=1.13 [1.09,1.16]$ and $c_c=.294[.293,.294]$. Data points represent at least $\geq10$ DR, with most representing $>30$DR. The coloring groups data points with identical $c$ values. The upper branch describes decreasing localization lengths (and thus insulating behavior) while the lower branch describes increasing localization lengths (and therefore metallic behavior)}
    \label{fig:scalingcurve}
\end{figure}

The quality of the fit was controlled by minimizing the relevant $\Bar{\chi}^2$ statistic (equation \ref{eq:lambdasq}), where $F_i$ is the value of the finite-size scaling function evaluated at the parameters used at the $i$th run of the TMM; $\Gamma_i$ is the computed value of $\Gamma$ in the TMM, and N is the total number of data points \cite{Slevin2014}.
\begin{equation}\label{eq:lambdasq}
    \Bar{\chi}^2=\sum_i^N(F_i-\Gamma_i)^2/\Gamma_i
\end{equation}
We minimized $\Bar{\chi}^2$ using the covariance matrix adaptation evolution strategy (CMA-ES) numerical optimization algorithm implemented in the parallel optimization library \verb|pagmo| \cite{Biscani2020}. We implemented bootstrap resampling to determine 95\% confidence intervals. This required sampling each dataset $\xi(E,W)$ $N$ times, with replacement, to create a bootstrapped data set with an identical size to the original dataset \cite{Efron1994-zs}. For each dataset, roughly 64 bootstrapped datasets were created and fed through the FSS protocol to generate a distribution of critical parameters which we used to estimate confidence intervals.
From our transfer matrix calculations, we obtained scaling curves using a wide range $c$ values, but we only performed the final FSS on values found within $10\%$ of the $c_c$ value. We chose logarithmically spaced $c$ values to capture scaling behavior over multiple orders of magnitude on either side of criticality. With this approach, we captured both the metallic $c>c_c$, and insulating $c<c_c$ regimes. Finally, from this fit, the critical exponent $\nu$ was determined as shown in figure \ref{fig:scalingcurve}. We have focused on single-parameter scaling in this work but a two-parameter scaling analysis may provide a more accurate value of $\nu$.

Next, we discuss the universal critical exponents of the just-identified MIT. For a review of known exponents of random Hamiltonian matrices and their universality classes, see \cite{Evers_2008, Chiu_2016}. There are two universality classes that are relevant to our system, the Orthogonal symmetry class (AI) with the critical exponent $\nu=1.57\pm0.003$ and Chiral Orthogonal symmetry class (BDI) with $\nu=1.12\pm0.06$ \cite{Wang_2021,Slevin2014} For a real and spinless system, orthogonal symmetry is defined as having time reversal symmetry (TRS), while Chiral symmetry comes either from having both particle-hole symmetry (PHS) and TRS or neither symmetry. 

Our Hamiltonian is real and spinless and thus has TRS, landing it in the Orthogonal symmetry class in the absence of any additional symmetries. Pure random hopping is known to belong to symmetry class BDI, which has an additional PHS, and thus, chiral symmetry \cite{Foster_2006,K_nig_2012}. However, our Hamiltonian only has PHS when $E=0$ and $W=0$, where the diagonal elements are then $0$. Any perturbation away from this, either in energy or on-site disorder, should break PHS, which would place our system in the AI symmetry class.

\section{Results}
\begin{figure}[hbt!]
    \centering
    \includegraphics[width=0.5\linewidth]{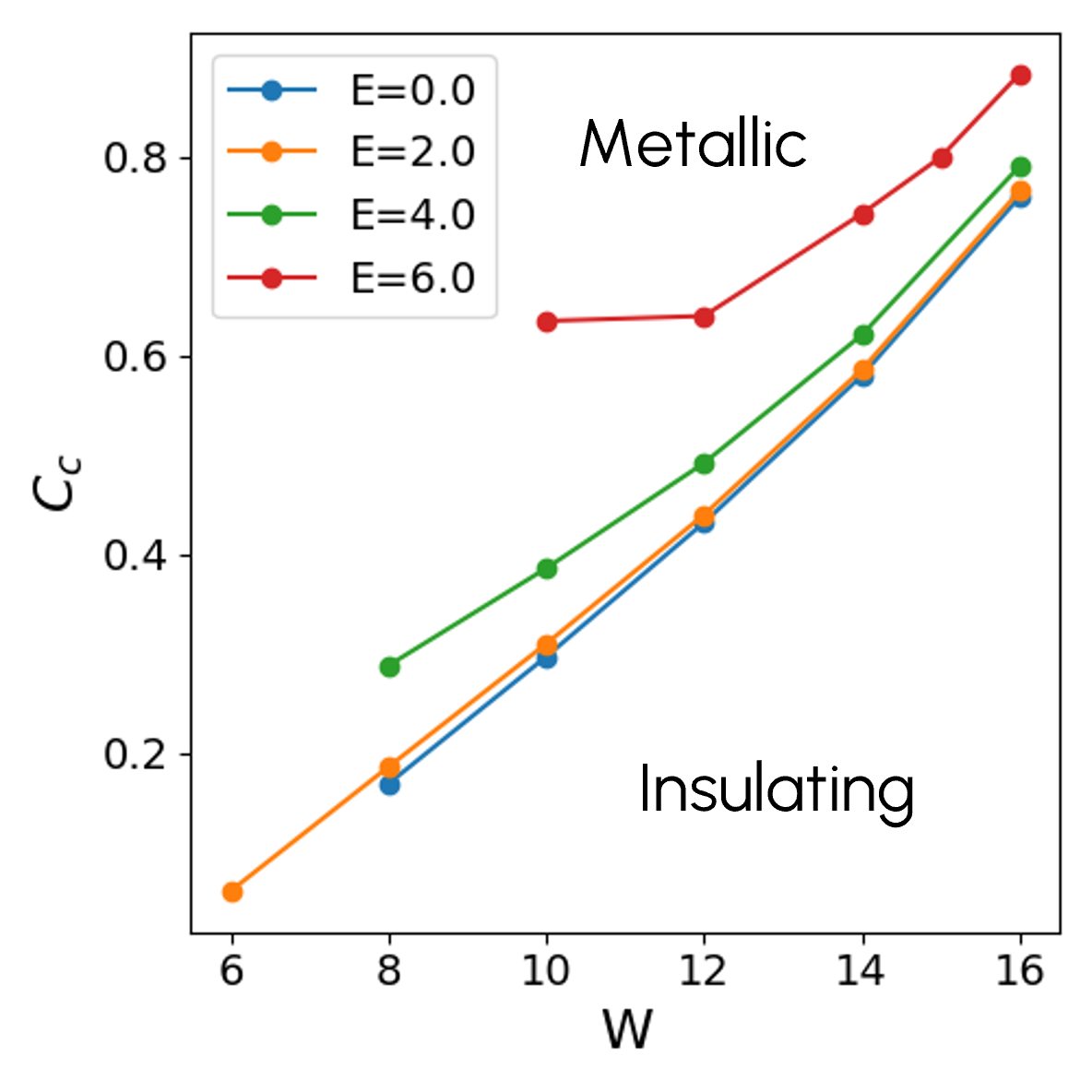}
    \caption{This phase diagram shows the critical concentration of high connectivity linkages (kinetic disorder) vs. the on-site disorder. Note that as the traditional Anderson critical on-site disorder ($W=16.53$) is approached, the phase boundaries converge with $c_c$ approaching unity. As the Fermi energy $E$ is increased, the insulating phase grows and the phase boundary becomes largely independent of on-site disorder}
    \label{fig:phase-diagram}
\end{figure}

Computing the critical concentration ($c_c$) of $t_H$ couplings allowed us to create one of our primary results: the dynamic phase diagram of the MIT in the doubly-disordered parameter space, as seen in figure \ref{fig:phase-diagram}. The phase diagram maps out phase boundaries between metallic (upper) and insulating (lower) regions. This phase diagram reproduces the traditional Anderson critical disorder limit, as seen by the existence of an insulating phase at all values of $c$ as $W \rightarrow16.53$ \cite{Slevin2014}. We estimate that for our choice of $t_L=0.3$, the phase transition at $c=0$ should occur at $W=W_c*t_L=16.53*0.3=4.96$, which is approached for slabs with $E$ near the Fermi level. Additionally, the insulating phase begins to fill the space as the on-site disorder-dependent mobility edge is approached ($E\rightarrow 6.5-7.4$) \cite{Nikoli__2001}. Within traditional Anderson theory, the system should be delocalized at sub-critical on-site disorder when the hopping parameter $t_{ij}$ is set to unity and the energy is not close to the mobility edge ($E< 6$) \cite{Filoche2024}. However, our phase diagram shows that kinetic disorder expands the insulating phase into the traditionally delocalized regime, indicating that low on-site disorder (at $E<6$) is not necessarily enough to reach metallic behavior. Additionally, the MIT can be crossed either by greatly reducing on-site disorder, or through increasing the fraction of $t_H$ couplings. 
\begin{figure}[hbt!]
    \centering
    \includegraphics[width=1.0\linewidth]{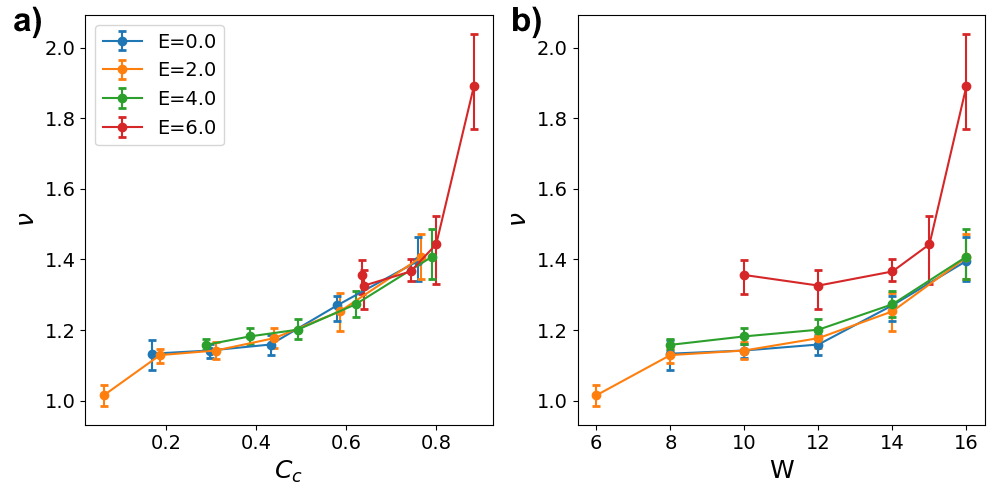}
    \caption{The critical exponent $\nu$ shown for a range of a) critical concentrations and b) disorders $W$, for 4 $E$ values. The crossover regime is seen as the plateau of critical values around $W=8-12$.}
    \label{fig:nuCW}
\end{figure}

We computed $\nu$ across our parameter space, as shown in figure \ref{fig:nuCW}. For most of the ($E,W$) space of our model, the critical exponent $\nu$ falls within the error of the BDI universality class, i.e. $\nu=~1.12\pm0.06$. This result is surprising, as it does not match up with a strict interpretation of the system's symmetries as any on-site disorder or finite energy should break PHS and reproduce the Anderson critical exponent of $1.57$. However, as $W$ and $E$ increase, we see the exponent flow towards the expected $\nu=1.57$ \cite{Slevin2014,Nikoli__2001}. The inclusion of kinetic disorder into the diagonal disorder Hamiltonian produces both unexpected critical exponents, and a MIT below the typical critical disorder $W_c$.

\begin{figure}
     \centering
     \includegraphics[width=0.65\linewidth]{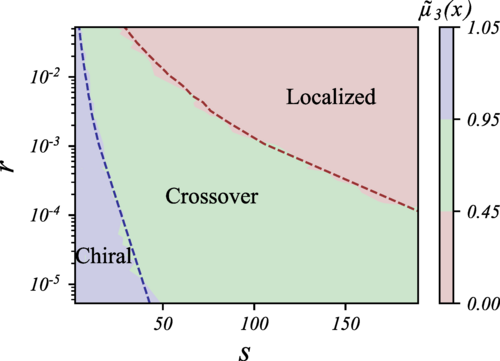}
     \caption{Evolution of the probability distribution $P(x)$ as a function of the two parameters $s=\textrm{Length}/\textrm{Mean Free Path}$ and $r=\textrm{Energy}*\textrm{Relaxation Time}$. In the blue region, for sufficiently small $s/|\ln r|^2$, the distribution is indistinguishable from that of the Chiral class. In the red region, when $s/|\ln r|^2$ is large, one enters the localized regime at any finite $r$, where the typical conductance decays exponentially with length. Here, the distribution $P(x)$ is approximately Gaussian, with a mean exceeding its standard deviation. In the green region, the distribution crosses over smoothly between these two limits. Reproduced from Kasturirangan et al. \cite{Kasturirangan_2022}}
     \label{fig:xover}
 \end{figure}

Our results suggest that there is a large intermediate region of low energy and low on-site disorder where PHS is not present but the critical exponent remains similar to that of the BDI class which includes PHS. This is a surprisingly extensive regime where the scaling is in the BDI regime with $\nu=1.12$, which is an experimentally observable quantity in 3D systems. Studies on the transition from class BDI to AI have been performed in 1D chains and metallic armchair graphene nanoribbons with hopping disorder using two parameter scaling, where Kasturirangan et al. found a similar crossover regime using DMPK analysis \cite{Kasturirangan_2022}. This crossover regime can be seen in figure \ref{fig:xover}. The reliability of our work is well-validated by the fact that the simulation reproduces the known exponent of $\nu=1.57$ of the Anderson transition when we induce disorder only on the diagonal. 

Another interesting result of our FSS analysis is that when the data includes values of $L<16$, there is significant degradation to the scaling curves. Surprisingly, the removal of the offending low $L$ values does not have a large effect on the final value of $\nu$. We believe the degradation is due to finite-size effects and when low $L$ values are removed, we find a noticeable difference in the convergence of the scaling curve. Reducing these finite-size effects in the transverse direction by increasing the highest $L$ calculated in our TMM was another key reason we sought to reduce $L_z$ and thus computation time. In general, we found using $L=17-30$ in our analysis made for the cleanest scaling. However, the fact that the calculated critical exponent is still within the margin of error for class BDI bolsters our confidence in the values our simulation has produced. 
 
\section{Conclusion}
This work investigated carrier transport in a model informed by experimental epitaxially-fused quantum dot solids, which were created and imaged by our collaborators Matt Law and Adam Moulé. Our goal was to identify pathways to cross the MIT, to achieve delocalization of electron wavefunctions, and to see mini-band formation in real samples. To this end, we have mapped out the dynamic phase diagram of these QD solids as seen in figure \ref{fig:phase-diagram}. We have made adjustments to the TMM and FSS analysis developed by Slevin, MacKinnon, and others, allowing us to increase our transverse system size up to $L=30$. We have shown that for our system, values of $L\leq16$ cause degradation in the FSS curve, which justifies our choice to reduce $L_z$ and increase $L$ past conventional values.

We found that the introduction of kinetic disorder drives the system through the MIT at values of on-site disorder below the traditional limit $W_c=16.5$. Kinetic disorder is a weak disorder however, having a maximum at $c=0.5$ in our implementation, which is insufficient to drive the system through the MIT at very low values of $E$ and $W$. For a large portion of the ($W$, $E$, $c$) parameter space, the critical exponent $\nu\approx 1.12$ was found to markedly differ from the purely diagonally-disordered case of $\nu\approx1.57$ of class AI, instead falling within error of the BDI class. Our results show that when there is weak or no on-site disorder our Hamiltonian falls in the BDI universality class, and with strong on-site disorder the system instead flows to class AI. 

This is strongly suggestive that the introduction of kinetic disorder into the weak Anderson diagonal disorder Hamiltonian changes the critical behavior around the MIT and that there is a cross-over regime between the universality classes of AI and BDI, at least for a binary distribution of highly conductive nearest-neighbor couplings. We note that in 3-dimensional systems the critical exponents corresponding to the divergence of the localization length $\nu$ and the critical exponent governing divergence of conductivity $s$ are the same \cite{Wegner_1976}, and thus $\nu$ is experimentally measurable. While real systems likely cannot be fabricated with both $W=0$ and $E=0$, our work shows that at sufficiently low values of both $E$ and $W$ a critical exponent of $\nu=s=1.12$ could be experimentally obtained.
\begin{acknowledgments}
We thank our invaluable collaborators Matt Law and Adam Moulé for their insight, analysis, and characterization efforts.

\end{acknowledgments}
\bibliographystyle{apsrev4-2.bst}
\bibliography{references}
\end{document}